\documentclass[twocolumn,english,aps,prl,superscriptaddress]{revtex4-1}
\usepackage[latin9]{inputenc}
\usepackage{color}
\usepackage{babel}
\usepackage{amsmath}
\usepackage{amssymb}
\usepackage{graphicx}
\usepackage[unicode=true,pdfusetitle,
 bookmarks=true,bookmarksnumbered=false,bookmarksopen=false,
 breaklinks=false,pdfborder={0 0 1},backref=false,colorlinks=true]
 {hyperref}
\hypersetup{
 urlcolor=blue, citecolor=blue, hyperfootnotes=blue, linkcolor=blue}
\usepackage{breakurl}

\makeatletter
\usepackage{mathrsfs}
\usepackage{epsfig}
\usepackage{amsfonts}
\usepackage[figuresright]{rotating}
\usepackage{dcolumn}
\usepackage{bm}
\usepackage{color}

\makeatother

\begin{document}

\title{Accessing strongly-coupled systems without compromising them}

\author{Xiangjin Kong}

\thanks{These authors contributed equally to this work.}

\affiliation{Department of Physics, National University of Defense Technology,
410073 Changsha, China}

\author{Carlos Navarrete-Benlloch}

\thanks{These authors contributed equally to this work.}

\affiliation{Wilczek Quantum Center, School of Physics and Astronomy, Shanghai
Jiao Tong University, Shanghai 200240, China}

\affiliation{Shanghai Research Center for Quantum Sciences, Shanghai 201315, China}

\author{Yue Chang}
\email{yuechang7@gmail.com}

\selectlanguage{english}%

\affiliation{Beijing Automation Control Equipment Institute, Beijing 100074, China}

\affiliation{Quantum Technology R$\&$D Center of China Aerospace Science and
Industry Corporation, Beijing 100074, China}
\begin{abstract}
The last decades have seen a burst of experimental platforms reaching
the so-called strong-coupling regime, where quantum coherent effects
dominate over incoherent processes such as dissipation and thermalization.
This has allowed us to create highly nontrivial quantum states and
put counterintuitive quantum-mechanical effects to test beyond the
wildest expectations of the founding fathers of quantum physics. The
strong-coupling regime comes with certain challenges though: the need
for a large isolation makes it difficult to access the system for
control or monitoring purposes. In this work we propose a way to access
such systems through an engineered environment that does not compromise
their strong-coupling effects. As a proof of principle, we apply the
approach to the photon-blockade effect present in nonlinear resonators,
but argue that the mechanism is quite universal. We also propose an
architecture based on superconducting circuits where the required
unconventional environment can be implemented, opening the way to
the experimental analysis of our ideas.
\end{abstract}
\maketitle
\textit{Introduction}.---The precise control of light-matter interactions
is arguably the landmark of quantum optical systems. It has allowed
us to handcraft quantum superposition states able to test the laws
of quantum mechanics well beyond what the founding fathers of quantum
mechanics ever thought possible. The generation of such states is
only possible by accessing the so-called strong-coupling regime, meaning
that the quantum-coherent part of the evolution induced by the interaction
between light and matter (governed by the Schrödinger equation) occurs
within a time scale where dissipation and other mechanisms responsible
for decoherence are still not relevant. Hence, since the pioneering
experiments in cavity quantum electrodynamics \citep{Raimond01,Miller05,Walther06}
and trapped ions \citep{Leibfried03,Schneider12}, people have worked
hard to achieve the strong-coupling regime in many other platforms,
including cold atoms \citep{Jaksch05,Bloch08}, superconducting circuits
\citep{Devoret13,Blais21}, and mechanical devices \citep{Aspelmeyer14}
among others.

Naturally, two strategies can be followed to achieve the strong-coupling
regime: enhancing the interactions or decreasing the decoherence rates.
One can try to enhance the light-matter interaction by confining the
system to smaller volumes or by finding matter with larger dipole
moments, but these are features that are not readily available in
most experimental platforms. The second route consists in isolating
the system incredibly well, but this comes at a high price: it becomes
then difficult to access it without compromising its strong coupling,
that is, to manipulate it and use its radiating fields for applications.

In this Letter we introduce a way to overcome this last limitation,
that is, of connecting the system to an environment that we can use
for driving and monitoring purposes, without compromising its strong-coupling
regime. Instead of a single environment, our idea uses two environments,
and an additional system (twin to the main one, but not necessarily
in the strong-coupling regime) that couples to these two environments
as well. When the systems couple symmetrically to the environments,
a perfectly isolated dark mode appears \citep{Lalumiere13}, which
inherits the interactions of the system under the right conditions.
By allowing for a slightly asymmetric coupling to the environments,
this mode acquires a small coupling into our controlled environment,
but effectively remaining in the strong-coupling regime. This mechanism
is universal, that is, works for any type of interaction. In order
to study it in detail, we consider the case of a nonlinear or Kerr
resonator \citep{Imamoglu97} and study the so-called photon blockade
effect, characteristic of the strong-coupling regime: from a weak
input coherent drive, we show that our mechanism indeed leads to strong
photon anti-bunching, while keeping large transmission probabilities,
and propose a specific architecture based on superconducting circuits
where our ideas can be explored.

\begin{figure*}[t]
\includegraphics[width=0.99\linewidth]{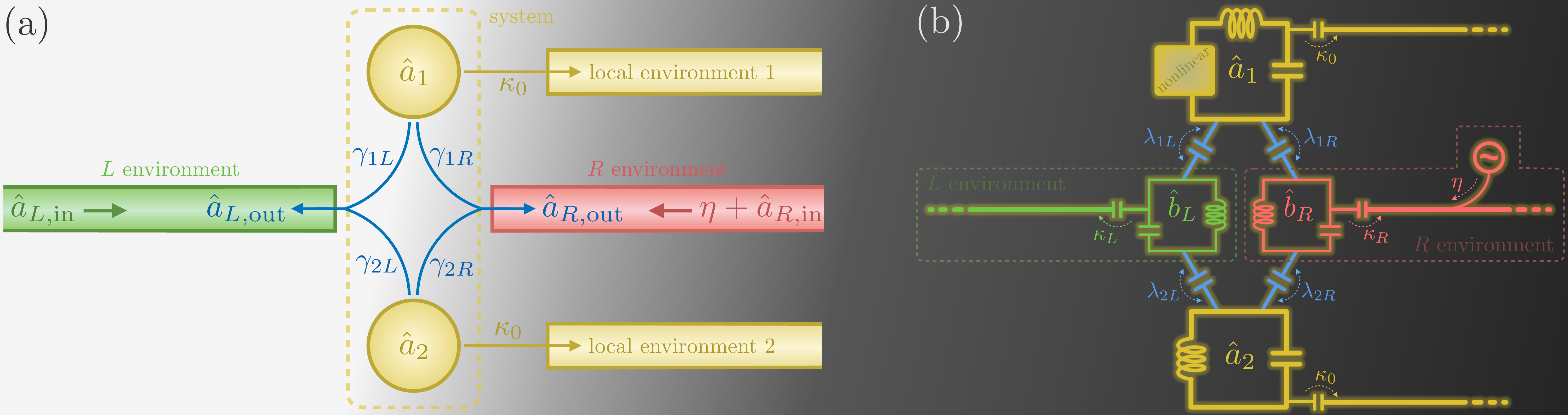} \caption{(Color online) (a) Sketch of our system, formed by two bosonic modes
$\{\hat{a}_{j}\}_{j=1,2}$, one of them presenting a nonlinearity
that is strong with respect to its local decay $\kappa_{0}$, but
weak with respect to the coupling to the $L$ and $R$ environments
that we use as input/output channels. The excitations of the bosonic
modes decay collectively to these two environments through outputs
$\{\hat{a}_{\alpha,\text{out}}\propto\sqrt{\gamma_{1\alpha}}\hat{a}_{1}+\sqrt{\gamma_{2\alpha}}\hat{a}_{2}\}_{\alpha=R,L}$,
while the environments are driven by input vacuum fluctuations $\{\hat{a}_{\alpha,\text{in}}\}_{\alpha=R,L}$
plus a coherent drive of irradiance $\eta^{2}$ for one environment.
As we explain in the text, an asymmetric decay of the modes into the
environments ($\gamma_{2R}\approx\gamma_{2L}\gg\gamma_{1R}=\gamma_{1L}$)
allows the system to reach an effective strong-coupling regime, as
evidenced by strong and robust anti-bunching of the output. (b) Sketch
of the experimental proposal. Two LC superconducting circuits (one
including some nonlinear element) form the main system, and exchange
excitations at rates $\lambda_{j\alpha}$ with two auxiliary LC circuits
with corresponding operators $\{\hat{b}_{\alpha}\}_{\alpha=R,L}$,
which in turn decay to two transmission lines at rates $\kappa_{\alpha}\gg\lambda_{j\alpha}$.
The same idea can be implemented in the optical domain, for example
by replacing the LC circuits with whispering gallery mode resonators
or photonic crystal cavities, and the transmission lines by photonic
fibers.}
\label{fig1} 
\end{figure*}

Note that our mechanism is completely different from the so-called
unconventional photon blockade effect \citep{Liew10,Bamba11,Flayac15,Flayac16},
which allows generating anti-bunched photon statistics even within
the weak-coupling regime by exploiting the coherent tunneling of quanta
within two resonators, but requires fine-tuning of the nonlinearity.

\emph{Model}.---We consider the open quantum system sketched in Fig.
\ref{fig1}(a), consisting of two bosonic modes described by annihilation
operators $\{\hat{a}_{j}\}_{j=1,2}$ obeying canonical commutation
relations $[\hat{a}_{j},\hat{a}_{l}^{\dagger}]=\delta_{jl}$ and $[\hat{a}_{j},\hat{a}_{l}]=0$.
Each mode couples to their local environments (inaccessible to the
experiment), which induce damping at rates $\kappa_{0}$. In addition,
the system couples to two experimentally accessible environments denoted
by $L$ and $R$ via the collective jump operators $\{\hat{J}_{\alpha}=\sqrt{\gamma_{1\alpha}}\hat{a}_{1}+\sqrt{\gamma_{2\alpha}}\hat{a}_{2}\}_{\alpha=R,L}$.
Hence, rather than considering only independent decay channels for
each bosonic mode, we consider a collective coherent decay that will
allow us to exploit quantum interference effects, especially when
allowing the four decay rates $\gamma_{j\alpha}$ to be independently
tunable. We later put forward concrete architectures based on currently
available superconducting-circuit technologies where our ideas should
be readily implementable, see Fig. \ref{fig1}(b).

Under standard Born-Markov conditions \citep{GardinerZollerBook,deVega17,CNB-QOnotes},
the fields coming out of the system into the collective environments
are characterized by output operators
\begin{equation}
\hat{a}_{\alpha,\mathrm{out}}(t)=\eta\delta_{\alpha,R}+\hat{a}_{\alpha,\mathrm{in}}(t)-\sqrt{2}\mathrm{i}\hat{J}_{\alpha}(t),
\end{equation}
where we have assumed that the system is driven from the $R$ environment
with flux $\eta^{2}$ (quanta per unit time), and $\hat{a}_{\alpha,\mathrm{in}}(t)$
annihilates the vacuum state of the environmental modes. Both the
input and output operators satisfy canonical commutation relations
in time, e.g. $[\hat{a}_{\alpha,\mathrm{in}}(t),\hat{a}_{\alpha',\mathrm{in}}^{\dagger}(t')]=\delta_{\alpha\alpha'}\delta(t-t')$
and $[\hat{a}_{\alpha,\mathrm{in}}(t),\hat{a}_{\alpha',\mathrm{in}}(t')]=0$.
We work in a picture rotating at the driving frequency, so the operators
in this expression are slowly-varying operators, not Heisenberg-picture
ones.

For the internal dynamics of the system, we consider linear modes
(harmonic oscillators) of the same frequency, one of them containing
a weak anharmonicity or Kerr nonlinearity (later we comment on other
types of nonlinearity). Again under Born-Markov conditions \citep{GardinerZollerBook,deVega17,CNB-QOnotes}
and working in a picture rotating at the driving frequency, the evolution
of the state of the system $\hat{\rho}$ is governed by the master
equation
\begin{equation}
\partial_{t}\hat{\rho}=-\mathrm{i}\left[\hat{H},\hat{\rho}\right]+\sum_{\alpha=L,R}\mathcal{D}_{J_{\alpha}}[\hat{\rho}]+\sum_{j=1}^{2}\kappa_{0}\mathcal{D}_{a_{j}}[\hat{\rho}],\label{MasterEq}
\end{equation}
where we have defined the Lindblad dissipator
\begin{equation}
\mathcal{D}_{J}[\hat{\rho}]=2\hat{J}\hat{\rho}\hat{J}^{\dagger}-\hat{J}^{\dagger}\hat{J}\hat{\rho}-\hat{\rho}\hat{J}^{\dagger}\hat{J},
\end{equation}
and the Hamiltonian (divided by $\hbar$) reads
\begin{equation}
\hat{H}=\Delta(\hat{a}_{1}^{\dagger}\hat{a}_{1}+\hat{a}_{2}^{\dagger}\hat{a}_{2})+g\hat{a}_{1}^{\dagger2}\hat{a}_{1}^{2}+\sqrt{2}\eta(\hat{J}_{R}+\hat{J}_{R}^{\dagger}),\label{HS}
\end{equation}
being $\Delta$ the detuning between the drive and the oscillators'
frequency.

We consider the situation in which strong coupling has been achieved
with respect to the local dissipative channels, while the coupling
remains weak with respect to the engineered environments used as input/output
channels, that is, $\kappa_{0}<g\ll\gamma_{j\alpha}$. In this work
we show how the nonlinear effects that are characteristic of the strong-coupling
regime can still be observed by exploiting quantum interference effects
coming from the collective character of the jump operators $\hat{J}_{\alpha}$,
allowing us to access the system without compromising its strong-coupling
regime.

\begin{figure*}[t]
\includegraphics[width=0.99\linewidth]{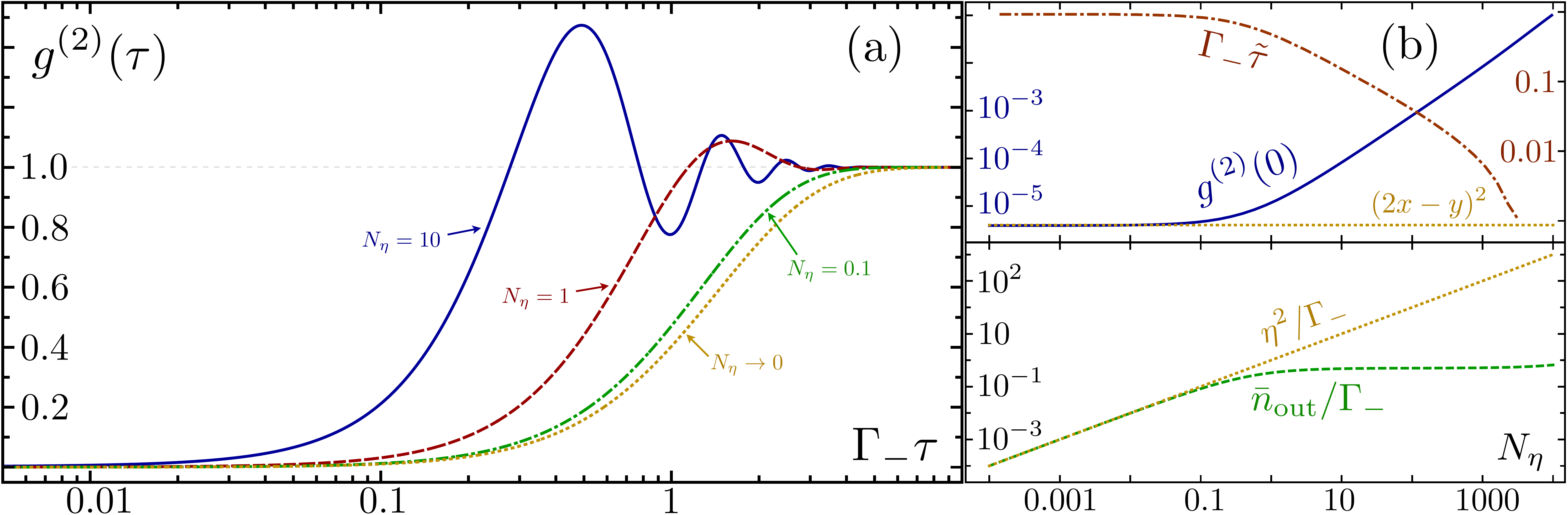}
\caption{(Color online) (a) Second-order correlation function $g^{(2)}(\tau)$
as a function of the normalized time delay $\Gamma_{-}\tau$ for different
values of the normalized input photon flux $N_{\eta}=2\Gamma_{-R}\eta^{2}/\Gamma_{-}^{2}$.
Strong anti-bunching is appreciated, even when $N_{\eta}>1$, where
our analytical results based on scattering theory (dotted yellow line)
cease to apply. (b) characterizes the anti-bunching as a function
of $N_{\eta}$, showing $g^{(2)}(0)$ in solid blue, the normalized
output photon flux $\bar{n}_{\text{out}}/\Gamma_{-}$ in dashed green,
and in dashed-dotted green the normalized time-delay $\Gamma_{-}\tilde{\tau}$
that it takes for the second-order correlation function to grow up
to $g^{(2)}(\tilde{\tau})=1/e$, providing a measure for the anti-bunching
time-scale. The dotted yellow curves are the analytic results obtained
from scattering theory. Note that on the upper panel the left ticks
refer to $g^{(2)}(0)$ while the right ones to $\Gamma_{-}\tilde{\tau}$.
In all plots we have chosen $y=0.01$, $g=4\times10^{-4}\gamma_{2}$,
$x=0.02\sqrt{g/\gamma_{1}}=0.004$, and $\kappa_{0}=0$.}
\label{fig2} 
\end{figure*}

\emph{Intuitive picture}.---In order to understand the essence of
the mechanism, it is best to consider the $\kappa_{0}=0$ case, and
perform a canonical transformation to the basis of modes that diagonalize
the dissipative part of the master equation:\begin{subequations}
\begin{align}
\hat{A}_{+} & =\hat{a}_{1}\cos\theta+\hat{a}_{2}\sin\theta,\\
\hat{A}_{-} & =\hat{a}_{1}\sin\theta-\hat{a}_{2}\cos\theta,
\end{align}
\end{subequations}with $\theta=\arctan[2\gamma_{12}/(\gamma_{1}-\gamma_{2}+\sqrt{(\gamma_{1}-\gamma_{2})^{2}+4\gamma_{12}^{2}})]\in[0,\pi/2]$,
where $\gamma_{j}=\gamma_{jR}+\gamma_{jL}$ and $\gamma_{12}=\sqrt{\gamma_{1R}\gamma_{2R}}+\sqrt{\gamma_{1L}\gamma_{2L}}$.
In terms of these new `normal' modes, which obey canonical commutation
relations unlike the jump operators $\hat{J}_{\alpha}$, the dissipative
part of (\ref{MasterEq}) takes the form
\begin{equation}
\sum_{\alpha=L,R}\mathcal{D}_{J_{\alpha}}[\hat{\rho}]=\sum_{n=\pm}\Gamma_{n}\mathcal{D}_{A_{n}}[\hat{\rho}],
\end{equation}
with normal damping rates $2\Gamma_{\pm}=\gamma_{1}+\gamma_{2}\pm\sqrt{(\gamma_{1}-\gamma_{2})^{2}+4\gamma_{12}^{2}}.$

Interestingly, when $\gamma_{jR}=\gamma_{jL}$ (that is, when the
modes couple in the same way to both environments), mode $\hat{A}_{+}$
gathers all the damping, $\Gamma_{+}=\gamma_{1}+\gamma_{2}$, while
mode $\hat{A}_{-}$ effectively becomes dark or undamped, $\Gamma_{-}=0$.
If, in addition, we work in the $\gamma_{2}\gg\gamma_{1}$ regime,
$\theta\approx\pi/2$, and the nonlinear term $g\hat{a}_{1}^{\dagger2}\hat{a}_{1}^{2}$
is dominated by the $g_{-}\hat{A}_{-}^{\dagger2}\hat{A}_{-}^{2}$
contribution, with $g_{-}=g\sin^{4}\theta$. Hence, while mode $\hat{A}_{+}$
is still in the weak-coupling regime, $\Gamma_{+}\gg g$, this is
not the case for mode $\hat{A}_{-}$, which in this extreme limit
behaves as a dissipativeless anharmonic or nonlinear oscillator.

This provides the intuition that by working close to this regime,
specifically keeping $y\equiv\gamma_{1}/\gamma_{2}\ll1$, but allowing
for a slightly asymmetric coupling of the modes to the environments,
mode $\hat{A}_{-}$ will effectively enter the strong-coupling regime
with $g_{-}\gg\Gamma_{-}$. Specifically, in the following we keep
mode-1's couplings symmetric for simplicity, $\gamma_{1R}=\gamma_{1L}=\gamma_{1}/2$,
and define the asymmetry parameter $x\equiv(\gamma_{2R}-\gamma_{2L})/\gamma_{2}$
for the second mode, with $|x|\ll1$. To the leading orders in $x$
and $y$, the $g_{-}\gg\Gamma_{-}$ condition can be recasted as $|x|\ll2\sqrt{g/\gamma_{1}}$,
which sets an upper bound to the asymmetry $x$ that we can introduce.

In the following, it is convenient to define $\sqrt{\Gamma_{+\alpha}}=\sqrt{\gamma_{1\alpha}}\cos\theta+\sqrt{\gamma_{2\alpha}}\sin\theta$
and $\sqrt{\Gamma_{-\alpha}}=\sqrt{\gamma_{1\alpha}}\sin\theta-\sqrt{\gamma_{2\alpha}}\cos\theta$,
which provide the relation between the collective jump operators and
the normal modes, $\hat{J}_{\alpha}=\sqrt{\Gamma_{+\alpha}}\hat{A}_{+}+\sqrt{\Gamma_{-\alpha}}\hat{A}_{-}$.

\emph{Photon blockade effect}.---In order to show that the intuitive
picture offered above is right, and effects of the nonlinearity are
observable through the collective environment, we consider now the
properties of the output field, in particular showing that it provides
antibunched quanta statistics, while allowing for near-perfect transmission
of the incident power. We gain analytical insight by considering the
weak-driving regime, performing an expansion in powers of $\eta$
\citep{SupMat}, equivalent to scattering theory formalism \citep{Munoz16,Caneva15,Shi15}.
The leading order of this expansion should be accurate as long as
the number of excitations in the system is small. Neglecting nonlinear
effects, the excitation number is dominated by $N_{\eta}\equiv2\Gamma_{-R}\eta^{2}/\Gamma_{-}^{2}$,
that is, the ratio between the driving and damping of the nearly-dark
mode $\hat{A}_{-}$, so the leading order of scattering theory is
expected to work as long as $N_{\eta}\ll1$. In the following we then
use $N_{\eta}$ instead of $\eta$ for our analysis.

Consider first the transmission amplitude, which to the lowest order
in $\eta$ we show in \citep{SupMat} to read
\begin{align}
\mathcal{T} & =\lim_{t\rightarrow\infty}\frac{\langle\hat{a}_{R,\mathrm{out}}(t)\rangle}{\eta}\approx1+2\mathrm{i}\sum_{n=\pm}\frac{\Gamma_{nR}}{\Delta-\mathrm{i}(\Gamma_{n}+\kappa_{0})}.\label{TransmissionAmplitude}
\end{align}
Using the definitions of the different rates, it is easy to show that
for a resonant drive ($\Delta=0$), we obtain $\mathcal{T}=(\Lambda x-1)/(\Lambda+1)$,
with $\Lambda=\kappa_{0}/\Gamma_{-}\approx4\kappa_{0}/\gamma_{2}x^{2}y$.
Hence, in the absence of local damping ($\Lambda=0$), all the energy
fed into the system in the form of an input coherent field is transferred
back to the $R$ environment, since $\mathcal{T}=-1$. However, the
quantum statistics of the field fed back into the environment are
dramatically changed by the interaction with the system. In particular,
we show in \citep{SupMat} that, again to the lowest order in $\eta$,
the normalized second-order correlation function takes the form
\begin{align}
 & g^{(2)}(\tau)\equiv\lim_{t\rightarrow\infty}\frac{\langle\hat{a}_{R,\mathrm{out}}^{\dagger}(t)\hat{n}_{R,\mathrm{out}}(t+\tau)\hat{a}_{R,\mathrm{out}}(t)\rangle}{\langle\hat{n}_{R,\mathrm{out}}(t)\rangle^{2}}\label{g2}\\
 & \approx\left[\left(\frac{1-\Lambda x}{1+\Lambda}\right)^{2}-\frac{y}{1+\Lambda}e^{-\tilde{\Gamma}_{+}\tau}-\frac{1-2x}{(1+\Lambda)^{2}}e^{-\tilde{\Gamma}_{-}\tau}\right]^{2},\nonumber 
\end{align}
where $\hat{n}_{R,\mathrm{out}}(t)=\hat{a}_{R,\mathrm{out}}^{\dagger}(t)\hat{a}_{R,\mathrm{out}}(t)$,
$\tilde{\Gamma}_{n}=\Gamma_{n}+\kappa_{0}$, and we have expanded
the final expression to first order in $x$ and $y$, assuming $\Gamma_{-}/g_{-}\ll1$
as well. For $\kappa_{0}=0$, this expression shows that $g^{(2)}(0)=(2x-y)^{2}$,
signaling strong anti-bunching over a time-scale determined by $\Gamma_{-}^{-1}$.
Moreover, the expression is independent of the coupling $g$, and
hence, such nonlinear effect does not require any fine-tuning of the
parameters, unlike previously proposed mechanisms \citep{Liew10,Bamba11,Flayac15,Flayac16}.

In Fig. \ref{fig2} we confront this analytical approximation with
the exact result found numerically \citep{CNB-NumericsNotes} from
the master equation (\ref{MasterEq}) with $\kappa_{0}=0$. In particular,
$g^{(2)}(\tau)$ is shown as a function of $\Gamma_{-}\tau$ in Fig.
\ref{fig2}(a) for different values of $N_{\eta}$. As expected, we
find good agreement when $N_{\eta}$ is small. Remarkably, even for
large $N_{\eta}$ anti-bunching is still present. In particular, we
characterize the anti-bunching as a function of $N_{\eta}$ in Figs.
\ref{fig2}(b) and (c), where we plot $g^{(2)}(0)$, the smallest
time-delay $\tilde{\tau}$ for which $g^{(2)}(\tilde{\tau})=1/e$
(characterizing the anti-bunching time-scale), and the output photon
flux $\bar{n}_{\text{out}}=\lim_{t\rightarrow\infty}\langle\hat{n}_{R,\mathrm{out}}(t)\rangle$.
As a single-photon source, we see that our system is optimized for
$N_{\eta}\approx1$, with $\bar{n}_{\text{out}}\approx\Gamma_{-}/2$,
$g^{(2)}(0)\approx10^{-5}$, and $\tilde{\tau}\approx\Gamma_{-}^{-1}$.

Note that in the case of non-vanishing local damping $\kappa_{0}\neq0$,
it is best to set the parameters such that $\Lambda$ is not very
large ($\Gamma_{-}\sim\kappa_{0}$), so that the transmission $\mathcal{T}$
remains reasonably large, while still achieving strong anti-bunching
$g^{(2)}(0)\ll1$. Below we show that this and all the previous conditions
are feasible with superconducting-circuit architectures.

\emph{Universality of the mechanism}.---Importantly, we have checked
that the photon blockade found in our system is not specific to the
Kerr nonlinearity. In particular, we have considered more intricate
nonlinearities such as coupling mode $\hat{a}_{1}$ to a two-level
system (Jaynes-Cummings model \citep{Tian92,Birnbaum05}) or to a
mechanical oscillator \citep{Rabl11}, finding that the output through
our collective decay channel presents the photon blockade effects
expected for such systems.

\textit{Implementation.}---We propose now a generic way in which
the collective decay required in our setup can be implemented, which
we support with specific parameters of current experimental platforms.
The basic idea is sketched in Fig.\ \ref{fig1}(b) through a superconducting-circuit
architecture. The original modes of our model, $\hat{a}_{j}$, correspond
to LC circuits, where the first one has an additional additional nonlinear
element (e.g., a Josephson junction for Ker nonlinearity \citep{Blais21},
but Jaynes-Cummings \citep{Blais21} or optomechanical \citep{Aspelmeyer14,Teufel11}
nonlinearities are also available in such platform). These circuits
are coupled to two identical auxiliary LC circuits with annihilation
operators $\{\hat{b}_{\alpha}\}_{\alpha=R,L}$, which in turn decay
each to a transmission line at rate $\{\kappa_{\alpha}\}_{\alpha=R,L}$.
In addition, the auxiliary circuit $R$ is resonantly driven with
an external generator. In a picture where all modes rotate at the
driving frequency, the master equation describing the dynamics of
the state of the whole system, denoted by $\hat{\rho}$, reads
\begin{equation}
\partial_{t}\hat{\rho}=-\mathrm{i}[\hat{H},\hat{\rho}]+\sum_{\alpha=R,L}\kappa_{\alpha}\mathcal{D}_{b_{\alpha}}[\hat{\rho}]+\sum_{j=1}^{2}\kappa_{0}\mathcal{D}_{a_{j}}[\hat{\rho}],\label{MasterEqAux}
\end{equation}
with
\begin{align}
\hat{H} & =\sum_{j=1}^{2}\Delta\hat{a}_{j}^{\dagger}\hat{a}_{j}+\hat{H}_{\mathrm{NL}}+\mathrm{i}\sqrt{2\kappa_{R}}\eta(\hat{b}_{R}^{\dagger}-\hat{b}_{R})\\
 & +\sum_{j=1}^{2}\sum_{\alpha=R,L}\lambda_{j\alpha}(\hat{a}_{j}\hat{b}_{\alpha}^{\dagger}+\hat{a}_{j}^{\dagger}\hat{b}_{\alpha}),\nonumber 
\end{align}
where $\hat{H}_{\mathrm{NL}}$ includes the nonlinear processes in
mode $\hat{a}_{1}$. When $\kappa_{\alpha}\gg\lambda_{j\alpha}$,
the auxiliary modes can be adiabatically eliminated as we show in
\citep{SupMat}, leading precisely to our original master equation
(\ref{MasterEq}), with $\gamma_{j\alpha}=\lambda_{j\alpha}^{2}/\kappa_{\alpha}$.
Hence, we see that the lossy auxiliary modes act as intermediate sinks
where the main modes interfere before decaying to the transmission
lines.

In order to prove the feasibility of our proposal with current experimental
platforms, let us consider some specific parameters, but keeping in
mind that our ideas do not require fine-tuning of these. We consider
LC circuits with $4$ GHz resonance frequency (all frequencies are
in $2\pi$ units in the following), and consider low-$Q$ auxiliary
circuits with $\kappa_{\alpha}=200$ MHz decay rates, and couplings
$\lambda_{1\alpha}=2$ MHz, $\lambda_{2R}=20.88$ MHz, and $\lambda_{2L}=19.08$
MHz. These values are common in this platform \citep{Blais21}, and
lead to $y\approx0.01$, $x\approx0.05$, and $\gamma_{2}\approx4$
MHz. Assuming then a state-of-the art $10^{8}$ quality factor for
the LC circuits, we obtain $\kappa_{0}=40$ Hz, while the nonlinearity
$g$ can easily reach KHz in these platform, so that we stay in the
$\kappa_{0}<g\ll\gamma_{j\alpha}$ regime. For this parameters, our
theory predicts then strong anti-bunching $g^{(2)}(0)\approx10^{-3}$
with large transmission $|\mathcal{T}|\approx0.64$.

\textit{Conclusions and outlook}.---In this work we have presented
a way to externally access systems that are already in the strong-coupling
regime, without compromising them. The idea relies on the addition
of an auxiliary system, which together with the main system is coupled
asymmetrically to two environments in an unconventional way: they
decay collectively, rather than independently. Using as an example
the photon-blockade effect of a Kerr resonator, we have shown that
the mechanism works, and agued that it is universal, in the sense
that it works for any type of nonlinearity. We have finally offered
a generic way in which the unconventional decays can be engineered,
showing that superconducting-circuit technologies are specially suited
for the observation of the phenomena predicted in this work.

\begin{acknowledgements} \textit{Acknowledgments}.---We thank Tao
Shi for useful suggestions. XJK acknowledges support by the National
Natural Science Foundation of China (NSFC) under Grant No. 11904404.
CNB acknowledges additional support from a Shanghai talent program
and Shanghai Municipal Science and Technology Major Project (Grant
No. 2019SHZDZX01). \end{acknowledgements}

\bibliographystyle{apsrev4-1}
\bibliography{EffectiveStrongCouplingBIB}

\newpage{}
\begin{widetext}
\begin{center}
\textbf{\Large{}Supplemental material}{\Large\par}
\par\end{center}
In this supplemental material we first explain how we have obtain
the analytical expressions for the transmission amplitude and the
two-time correlation function in the weak driving limit. Next we show
how the adiabatic elimination of the auxiliary modes leads to the
model that we seek with the required unconventional collective decays.
\begin{center}
\textbf{\large{}I. Weak-driving analytics}{\large\par}
\par\end{center}
Consider the master equation (\ref{MasterEq}) presented in the main
text, which we rewrite as
\begin{equation}
\partial_{t}\hat{\rho}=\mathcal{L}\hat{\rho}=\underbrace{\mathcal{L}_{\mathrm{eff}}[\hat{\rho}]+\mathcal{L}_{\mathrm{jump}}[\hat{\rho}]}_{\mathcal{L}_{0}[\hat{\rho}]}+\mathcal{L}_{D}[\hat{\rho}],
\end{equation}
where
\begin{equation}
\mathcal{L}_{\mathrm{jump}}[\hat{\rho}]=\sum_{n=\pm}\Gamma_{n}\hat{A}_{n}\hat{\rho}\hat{A}_{n}^{\dagger}+\gamma_{0}\sum_{j=1}^{2}\hat{a}_{j}\hat{\rho}\hat{a}_{j}^{\dagger},
\end{equation}
contains the terms responsible for irreversible quantum jumps, while
$\mathcal{L}_{\mathrm{eff}}[\hat{\rho}]=-\mathrm{i}[\hat{H}_{\mathrm{eff}},\hat{\rho}]$
and $\mathcal{L}_{D}[\hat{\rho}]=-\mathrm{i}[\hat{H}_{D},\hat{\rho}]$
contain the reversible part of the dynamics effected, respectively,
by the effective non-Hermitian and the driving Hamiltonians\begin{subequations}
\begin{align}
\hat{H}_{\mathrm{eff}} & =\hat{H}_{0}-\mathrm{i}\sum_{n=\pm}\Gamma_{n}\hat{A}_{n}^{\dagger}\hat{A}_{n}-\mathrm{i}\gamma_{0}\sum_{j=1}^{2}\hat{a}_{j}^{\dagger}\hat{a}_{j},\\
\hat{H}_{D} & =\sqrt{2}\eta(\hat{J}_{R}+\hat{J}_{R}^{\dagger})=\eta\left(\sqrt{2\Gamma_{+R}}\hat{A}_{+}+\sqrt{2\Gamma_{-R}}\hat{A}_{-}+\mathrm{H.c.}\right),
\end{align}
 \end{subequations}with system Hamiltonian 
\begin{equation}
\hat{H}_{0}=\sum_{n=\pm}\Delta\hat{A}_{n}^{\dagger}\hat{A}_{n}+g\left(\hat{A}_{+}^{\dagger}\cos\theta+\hat{A}_{-}^{\dagger}\sin\theta\right)^{2}\left(\hat{A}_{+}\cos\theta+\hat{A}_{-}\sin\theta\right)^{2}.
\end{equation}
Note that except for the local decay term, all other terms are written
in terms of the modes that diagonalize the collective dissipation.

Assuming a sufficiently week driving $\eta$, so that $\mathcal{L}_{D}$
acts just as a perturbation to $\mathcal{L}_{0}$, it is convenient
to write the time-evolution super-operator as the Dyson expansion
\citep{GardinerZollerBook}
\begin{equation}
e^{\mathcal{L}t}=\sum_{n=0}^{\infty}\int_{0}^{t}dt_{n}e^{\mathcal{L}_{0}(t-t_{n})}\mathcal{L}_{D}\int_{0}^{t_{n}}dt_{n-1}e^{\mathcal{L}_{0}(t_{n}-t_{n-1})}\mathcal{L}_{D}...\int_{0}^{t_{2}}dt_{1}e^{\mathcal{L}_{0}(t_{2}-t_{1})}\mathcal{L}_{D}e^{\mathcal{L}_{0}t_{1}},\label{DysonLD}
\end{equation}
which we will truncate at any desired order in $\mathcal{L}_{D}$.
Moreover, applying a similar expansion to the evolution induced by
$\mathcal{L}_{0}$,
\begin{equation}
e^{\mathcal{L}_{0}t}=\sum_{n=0}^{\infty}\int_{0}^{t}dt_{n}e^{\mathcal{L}_{\mathrm{eff}}(t-t_{n})}\mathcal{L}_{\mathrm{jump}}\int_{0}^{t_{n}}dt_{n-1}e^{\mathcal{L}_{\mathrm{eff}}(t_{n}-t_{n-1})}\mathcal{L}_{\mathrm{jump}}...\int_{0}^{t_{2}}dt_{1}e^{\mathcal{L}_{\mathrm{eff}}(t_{2}-t_{1})}\mathcal{L}_{\mathrm{jump}}e^{\mathcal{L}_{\mathrm{eff}}t_{1}},\label{DysonLjump}
\end{equation}
we see that for the vacuum state $|0\rangle$ and any operator $\hat{P}$
we have\begin{subequations}
\begin{align}
e^{\mathcal{L}_{0}\tau}[\hat{P}|0\rangle\langle0|] & =e^{-\mathrm{i}\hat{H}_{\mathrm{eff}}\tau}\hat{P}|0\rangle\langle0|,\\
e^{\mathcal{L}_{0}\tau}[|0\rangle\langle0|\hat{P}] & =|0\rangle\langle0|\hat{P}e^{\mathrm{i}\hat{H}_{\mathrm{eff}}^{\dagger}\tau},
\end{align}
\end{subequations}which is easily proven from\begin{subequations}
\begin{align}
e^{\mathcal{L}_{\mathrm{eff}}\tau}[\hat{B}] & =e^{-\mathrm{i}\hat{H}_{\mathrm{eff}}\tau}\hat{B}e^{\mathrm{i}\hat{H}_{\mathrm{eff}}^{\dagger}\tau},\\
\mathcal{L}_{\mathrm{jump}}[\hat{P}|0\rangle\langle0|] & =0=\mathcal{L}_{\mathrm{jump}}[|0\rangle\langle0|\hat{P}],
\end{align}
\end{subequations}for any operator $\hat{B}$. Using all these expressions,
we can evaluate the observables of interest to the lowest nontrivial
order in the driving $\eta$. In the case of the transmission amplitude
$\mathcal{T}$ of Eq. (\ref{TransmissionAmplitude}) in the main text,
it's enough to keep terms up to first order in $\eta$, so that we
can approximate the steady state of the system by
\begin{align}
\bar{\rho} & =\lim_{t\rightarrow\infty}e^{\mathcal{L}t}[|0\rangle\langle0|]\approx\lim_{t\rightarrow\infty}e^{\mathcal{L}_{0}t}[|0\rangle\langle0|]+\lim_{t\rightarrow\infty}\int_{0}^{t}dt_{1}e^{\mathcal{L}_{0}(t-t_{1})}\mathcal{L}_{D}e^{\mathcal{L}_{0}t_{1}}[|0\rangle\langle0|]\\
 & =|0\rangle\langle0|-\sqrt{2}\mathrm{i}\eta\lim_{t\rightarrow\infty}\int_{0}^{t}dt_{1}e^{\mathcal{L}_{0}(t-t_{1})}\left(\hat{J}_{R}^{\dagger}|0\rangle\langle0|-|0\rangle\langle0|\hat{J}_{R}\right)\nonumber \\
 & =|0\rangle\langle0|-\sqrt{2}\mathrm{i}\eta\lim_{t\rightarrow\infty}\int_{0}^{t}dt_{1}\left(e^{-\mathrm{i}\hat{H}_{\mathrm{eff}}(t-t_{1})}\hat{J}_{R}^{\dagger}|0\rangle\langle0|-|0\rangle\langle0|\hat{J}_{R}e^{\mathrm{i}\hat{H}_{\mathrm{eff}}^{\dagger}(t-t_{1})}\right)\nonumber \\
 & =|0\rangle\langle0|-\sqrt{2}\eta\left(\hat{H}_{\mathrm{eff}}^{-1}\hat{J}_{R}^{\dagger}|0\rangle\langle0|-|0\rangle\langle0|\hat{J}_{R}\hat{H}_{\mathrm{eff}}^{-1\dagger}\right).\nonumber 
\end{align}
where we start from the vacuum state for convenience, noting that
the steady state of the system is unique, and we assumed that the
imaginary parts of the eigenvalues of $\hat{H}_{\mathrm{eff}}$ are
all negative (except the one corresponding to the vacuum state, which
plays no role in the expression anyways), as corresponds to a physical
system. Thus, taking into account that $\langle\hat{a}_{\alpha,\mathrm{in}}(t)\rangle=0$
and $\hat{J}_{R}|0\rangle=0$, we obtain
\begin{equation}
\mathcal{T}\approx\frac{\eta-\sqrt{2}\mathrm{i}\mathrm{tr}\{\hat{J}_{R}\bar{\rho}\}}{\eta}=1+2\mathrm{i}\left\langle 0\right\vert \hat{J}_{R}\hat{H}_{\mathrm{eff}}^{-1}\hat{J}_{R}^{\dagger}\left\vert 0\right\rangle .\label{T_sup}
\end{equation}
Finally, noting that $\hat{H}_{\mathrm{eff}}$ conserves the number
of excitations, the final analytical expression (\ref{TransmissionAmplitude})
presented in the main text is straightforwardly obtained by representing
$\hat{H}_{\mathrm{eff}}$ in the single-photon subspace spanned by
$\{\hat{A}_{\pm}^{\dagger}|0\rangle\}$, obtaining a $2\times2$ matrix
that is easily inverted.

The two-time correlation function $g^{(2)}(\tau)$ of Eq. (\ref{g2})
in the main text is found in a similar way, just requiring lengthier
algebra that we skip here for brevity. In particular, the numerator
of the expression can be written as
\begin{align}
\lim_{t\rightarrow\infty}\langle\hat{a}_{R,\mathrm{out}}^{\dagger}(t)\hat{n}_{R,\mathrm{out}}(t+\tau)\hat{a}_{R,\mathrm{out}}(t)\rangle & =\lim_{t\rightarrow\infty}\langle\hat{K}^{\dagger}(t)\hat{K}^{\dagger}(t+\tau)\hat{K}(t+\tau)\hat{K}(t)\rangle=\mathrm{tr}\left\{ \hat{K}\left(e^{\mathcal{L}\tau}[\hat{K}\bar{\rho}\hat{K}^{\dagger}]\right)\hat{K}^{\dagger}\right\} ,
\end{align}
with $\hat{K}=\eta-\sqrt{2}\mathrm{i}\hat{J}_{R}$, where we have
removed the contribution from $\hat{a}_{R,\mathrm{in}}$ because the
causality relation $[\hat{a}_{R,\mathrm{in}}(t),\hat{J}_{R}(t'<t)]=0$
\citep{GardinerZollerBook} allows bringing all input annihilation
(creation) operators to the right (left) of the expression and destroy
the environmental vacuum state, and we have applied the quantum regression
theorem \citep{GardinerZollerBook} in the last step. Using now (\ref{DysonLD})
and (\ref{DysonLjump}), and keeping terms up to fourth order in $\eta$,
one can find after some careful algebra
\begin{equation}
\mathrm{tr}\left\{ \hat{K}\left(e^{\mathcal{L}\tau}[\hat{K}\bar{\rho}\hat{K}^{\dagger}]\right)\hat{K}^{\dagger}\right\} =\eta^{4}\left\vert w\left(\tau\right)\right\vert ^{2},
\end{equation}
where
\begin{equation}
w\left(\tau\right)=\mathcal{T}^{2}-2\left\langle 0\right\vert \hat{J}_{R}e^{-\mathrm{i}\hat{H}_{\mathrm{eff}}\tau}\hat{J}_{R}\hat{H}_{\mathrm{eff}}^{-1}\hat{J}_{R}^{\dagger}\hat{H}_{\mathrm{eff}}^{-1}\hat{J}_{R}^{\dagger}\left\vert 0\right\rangle +2\left\langle 0\right\vert \hat{J}_{R}e^{-\mathrm{i}\hat{H}_{\mathrm{eff}}\tau}\hat{H}_{\mathrm{eff}}^{-1}\hat{J}_{R}^{\dagger}\left\vert 0\right\rangle \left\langle 0\right\vert \hat{J}_{R}\hat{H}_{\mathrm{eff}}^{-1}\hat{J}_{R}^{\dagger}\left\vert 0\right\rangle ,
\end{equation}
is known as two-particle wave-function in the context of scattering
theory \citep{Munoz16}. Noting that to second order in $\eta$ the
steady-state photon number is given by
\begin{equation}
\lim_{t\rightarrow\infty}\langle\hat{n}_{R,\mathrm{out}}(t)\rangle\approx\eta^{2}\left\vert \mathcal{T}\right\vert ^{2},
\end{equation}
the second-order correlation function (\ref{g2}) is finally found
as
\begin{equation}
g^{\left(2\right)}\left(\tau\right)\approx\frac{\left\vert w\left(\tau\right)\right\vert ^{2}}{\left\vert \mathcal{T}\right\vert ^{4}}.\label{g2_Sup}
\end{equation}
Note that expressions (\ref{T_sup}) and (\ref{g2_Sup}) can also
be acquired from the scattering theory formalism \citep{Shi15} since
it has been proved that, in the weak-driving limit, scattering theory
is equivalent to the master equation when it comes to the evaluation
of observables \citep{Caneva15}. In this case, the final analytical
expression (\ref{g2}) provided in the main text is found by representing
$\hat{H}_{\mathrm{eff}}$ in the single- or two-photon subspaces spanned
by $\{\hat{A}_{\pm}^{\dagger}|0\rangle\}$ and $\{\hat{A}_{+}^{\dagger}\hat{A}_{-}^{\dagger}|0\rangle,\hat{A}_{\pm}^{\dagger2}|0\rangle/\sqrt{2}\}$,
respectively, as required in the corresponding expression, and doing
then a Taylor expansion in the small parameters $x$, $y$, and $\Gamma_{-}/g_{-}$
defined in the main text.
\begin{center}
\textbf{\large{}II. Elimination of the auxiliary modes}{\large\par}
\par\end{center}
Consider the master equation (\ref{MasterEqAux}) in the main text.
Here we show how the auxiliary modes can be eliminated, leading to
the desired model defined by master equation (\ref{MasterEq}) presented
in the main text for the main modes. In order to do so, we apply the
projection super-operator technique as presented in \citep{CNB-QOnotes,GardinerZollerBook}.
In particular, this approach considers the effect that the auxiliary
modes (dubbed `environment' in this context) have on the dynamics
of the system, assuming Born-Markov conditions to their interaction
(precisely defined below), which is shown to be a good approximation
as long as the decay rates $\kappa_{\alpha}$ of the auxiliary modes
dominates over any other rate affecting the system's dynamics, including
the couplings $\lambda_{j\alpha}$. Implicit in the Born-Markov approximation
is the absence of back-action from the system to the auxiliary modes,
such that from the point of view of the system dynamics, the auxiliary
modes remain in the coherent steady state that they would be in the
absence of interaction. It is then convenient to move to a displaced
picture where the coherent contribution to the state of the auxiliary
modes is removed, defined by a unitary displacement transformation
$\hat{D}(\beta)=\exp[\beta(\hat{b}_{R}^{\dagger}-\hat{b}_{R})]$ with
$\beta=\sqrt{2/\kappa_{R}}\eta$. The transformed state $\tilde{\rho}=\hat{D}^{\dagger}(\beta)\hat{\rho}\hat{D}(\beta)$
evolves then according to the master equation
\begin{equation}
\partial_{t}\tilde{\rho}=\mathcal{L}_{S}[\tilde{\rho}]+\mathcal{L}_{E}[\tilde{\rho}]+\mathcal{L}_{1}[\tilde{\rho}],\label{MasterEqAuxSup}
\end{equation}
with\begin{subequations}
\begin{align}
\mathcal{L}_{S}[\tilde{\rho}] & =-\mathrm{i}[\hat{H}_{S},\tilde{\rho}]+\sum_{j=1}^{2}\kappa_{0}\mathcal{D}_{a_{j}}[\tilde{\rho}],\\
\mathcal{L}_{E}[\tilde{\rho}] & =\sum_{\alpha=R,L}\kappa_{\alpha}\mathcal{D}_{b_{\alpha}}[\tilde{\rho}],\\
\mathcal{L}_{1}[\tilde{\rho}] & =-\mathrm{i}[\hat{H}_{1},\tilde{\rho}],
\end{align}
\end{subequations}and\begin{subequations}
\begin{align}
\hat{H}_{S} & =\sum_{j=1}^{2}\Delta\hat{a}_{j}^{\dagger}\hat{a}_{j}+\hat{H}_{\mathrm{NL}}+\beta(\hat{S}_{R}+\hat{S}_{R}^{\dagger}),\label{HS-sup}\\
\hat{H}_{1} & =\sum_{\alpha=R,L}(\hat{S}_{\alpha}\hat{b}_{\alpha}^{\dagger}+\hat{S}_{\alpha}^{\dagger}\hat{b}_{\alpha}),
\end{align}
\end{subequations}where $\hat{S}_{\alpha}=\sum_{j=1}^{2}\lambda_{j\alpha}\hat{a}_{j}$.
Comparing the Hamiltonians (\ref{HS-sup}) right above and (\ref{HS})
in the main text, we can already identify $\gamma_{jR}=\lambda_{jR}^{2}/\kappa_{R}$
for the parameters of the collective operator $\hat{J}_{R}=\sum_{j=1}^{2}\sqrt{\gamma_{1R}}\hat{a}_{j}$.
We then define the projection super-operator $\mathcal{P}[\hat{C}]=\mathrm{tr}_{E}\{\hat{C}\}\otimes\bar{\rho}_{E}$,
where $\hat{C}$ is an arbitrary operator acting on the full Hilbert
space, $\mathrm{tr}_{E}$ denotes the partial trace over the auxiliary
or environmental modes, whose reference $\bar{\rho}_{E}=|\text{vac}\rangle\langle\text{vac}|$
is the vacuum state, which is their steady state in this picture,
$\mathcal{L}_{E}[\bar{\rho}_{E}]=0$. The projection super-operator
$\mathcal{P}$ divides the operator space into relevant and irrelevant
sectors, the latter obtained from the complementary projector $\mathcal{Q}=1-\mathcal{P}$.
Projecting the master equation (\ref{MasterEqAuxSup}) onto these
super-operators, formally integrating the equation for $\mathcal{Q}[\tilde{\rho}]$,
substituting on the equation of $\mathcal{P}[\tilde{\rho}]$, and
keeping only terms up to quadratic order on the interaction $\hat{H}_{1}$
(so-called Born approximation), it is easy \citep{CNB-QOnotes} to
arrive to the following equation for the reduced state of the system
$\hat{\rho}_{S}=\mathrm{tr}_{E}\{\tilde{\rho}\}$,
\begin{equation}
\partial_{t}\hat{\rho}_{S}=\mathcal{L}_{S}[\hat{\rho}_{S}]+\sum_{\alpha=R,L}\int_{0}^{t}d\tau C_{\alpha}(\tau)\left(\hat{S}_{\alpha}e^{\mathcal{L}_{S}\tau}[\hat{\rho}_{S}(t-\tau)\hat{S}_{\alpha}^{\dagger}]-\hat{S}_{\alpha}^{\dagger}e^{\mathcal{L}_{S}\tau}[\hat{S}_{\alpha}\hat{\rho}_{S}(t-\tau)]+\mathrm{H.c.}\right),
\end{equation}
where we have defined the two-time correlators \citep{CNB-QOnotes}
\begin{align}
C_{\alpha}(\tau) & =\mathrm{tr}\left\{ \hat{b}_{\alpha}^{\dagger}e^{\mathcal{L}_{E}\tau}\left[\bar{\rho}_{E}\hat{b}_{\alpha}\right]\right\} =\lim_{t\rightarrow\infty}\left\langle \hat{b}_{\alpha}(t)\hat{b}_{\alpha}^{\dagger}(t+\tau)\right\rangle \\
 & =\mathrm{tr}\left\{ \hat{b}_{\alpha}e^{\mathcal{L}_{E}\tau}\left[\hat{b}_{\alpha}^{\dagger}\bar{\rho}_{E}\right]\right\} =\lim_{t\rightarrow\infty}\left\langle \hat{b}_{\alpha}(t+\tau)\hat{b}_{\alpha}^{\dagger}(t)\right\rangle =e^{-\kappa_{\alpha}\tau}.\nonumber 
\end{align}
where the equality between the first and second lines holds for this
specific case, but not in general \citep{CNB-QOnotes,GardinerZollerBook}.
The final step consists in assuming that $\kappa_{\alpha}$ is much
larger than any other scale in the system's dynamics, so that we can
neglect the $\tau$ dependence $e^{\mathcal{L}_{S}\tau}[\hat{\rho}_{S}(t-\tau)\hat{S}_{\alpha}^{\dagger}]$
and $e^{\mathcal{L}_{S}\tau}[\hat{S}_{\alpha}\hat{\rho}_{S}(t-\tau)]$,
so-called Markov approximation. Considering then times $t\gg\kappa_{\alpha}^{-1}$,
we obtain the desired master equation
\begin{equation}
\partial_{t}\hat{\rho}_{S}=\mathcal{L}_{S}[\hat{\rho}_{S}]+\sum_{\alpha=R,L}\mathcal{D}_{J_{\alpha}}[\hat{\rho}_{S}],
\end{equation}
with the identification $\gamma_{j\alpha}=\lambda_{j\alpha}^{2}/\kappa_{\alpha}$,
so that $\hat{J}_{\alpha}=\hat{S}_{\alpha}/\sqrt{\kappa_{\alpha}}$.
Note that the Markov approximation holds as long as $\kappa_{\alpha}$
dominates over $\gamma_{j\alpha}$, $\kappa_{0}$, $\Delta$, and
the rates of the nonlinear interaction, e.g., $g$ in the case of
Kerr nonlinearity.\newpage{}
\end{widetext}

\end{document}